\documentclass[conference]{IEEEtran}
\IEEEoverridecommandlockouts

\usepackage{hyperref}
\usepackage{cite}
\usepackage{amsmath,amssymb,amsfonts,amsthm, mathtools,stmaryrd}
\usepackage{graphicx}
\usepackage{textcomp}
\usepackage{xcolor}
\def\BibTeX{{\rm B\kern-.05em{\sc i\kern-.025em b}\kern-.08em
    T\kern-.1667em\lower.7ex\hbox{E}\kern-.125emX}}
\usepackage[capitalize]{cleveref}
\usepackage{bm}
\usepackage{braket}
\usepackage{xspace}
\usepackage{multirow}
\usepackage{nicefrac}
\usepackage{booktabs}
\usepackage{subcaption}

\usepackage{pgfplots}
\usepgfplotslibrary{external}
\crefname{algocf}{alg.}{algs.}
\crefname{algocf}{Alg.}{Algs.}
\usepackage[noend]{algpseudocode}
\usepackage[resetcount,ruled,noline,linesnumbered]{algorithm2e}

\newcommand{\bv}{\boldsymbol{v}}
\newcommand{\bW}{\boldsymbol{W}}
\newcommand{\bp}{\boldsymbol{p}}
\newcommand{\bP}{\boldsymbol{P}}
\newcommand{\bc}{\boldsymbol{c}}
\newcommand{\bs}{\boldsymbol{s}}
\newcommand{\bZ}{\boldsymbol{Z}}
\newcommand{\bS}{\boldsymbol{S}}
\newcommand{\by}{\boldsymbol{y}}
\newcommand{\bX}{\boldsymbol{X}}

\newcommand{\bM}{\boldsymbol{M}}
\newcommand{\bQ}{\boldsymbol{Q}}

\newcommand{\btau}{\boldsymbol{\tau}}

\newcommand{\ones}[1]{\boldsymbol{1}_{#1}}
\newcommand{\ident}[1]{\boldsymbol{I}_{#1}}

\newcommand{\ups}[1]{\boldsymbol{U}_{#1}}

\newcommand{\trans}{\intercal}

\newcommand{\indices}[1]{[#1]}
\newcommand{\iverson}[1]{\left\llbracket #1\, \right\rrbracket}

\newcommand{\bx}{\boldsymbol{x}}

\newcommand{\g}[1]{#1}

\newcommand{\ld}[1]{\hat{#1}}

\newcommand{\adjswitch}{\boldsymbol{A}}

\newcommand{\maxload}{\boldsymbol{M}}

\newcommand{\fmin}[1]{{#1^{\downarrow}}}
\newcommand{\fmax}[1]{{#1^{\uparrow}}}
\newcommand{\fmins}[1]{{#1^{*\downarrow}}}
\newcommand{\fmaxs}[1]{{#1^{*\uparrow}}}
\newcommand{\fnorm}[1]{\left\lceil #1 \right\rceil}

\newcommand{\cyclefunc}{F}
\newcommand{\swapmat}{\mat{R}}

\renewcommand{\vec}[1]{\bm{#1}}
\newcommand{\mat}[1]{\bm{#1}}

\newcommand*{\defeq}{\mathrel{\vcenter{\baselineskip0.5ex\lineskiplimit0pt\hbox{\scriptsize.}\hbox{\scriptsize.}}}%
	=}

\DeclareMathOperator*{\argmin}{arg\,min}

\DeclareMathOperator*{\diag}{diag}

\newcommand{\QUBO}{\textsc{QUBO}\@\xspace}

\begin{document}

\title{Multi-Objective Quantum Power System Redispatch}

\author{%
  \IEEEauthorblockN{
{Loong Kuan Lee}\IEEEauthorrefmark{1}\IEEEauthorrefmark{3},
{Thore Gerlach}\IEEEauthorrefmark{1},
{Johannes Knaute}\IEEEauthorrefmark{2},
{Florian Gerhardt}\IEEEauthorrefmark{2},
{Patrick V\"olker}\IEEEauthorrefmark{2},
{Tomislav Maras}\IEEEauthorrefmark{2},\\
{Alexander Dotterweich}\IEEEauthorrefmark{2} and
{Nico Piatkowski}\IEEEauthorrefmark{1}
}
\IEEEauthorblockA{
  \IEEEauthorrefmark{1}%
  Fraunhofer IAIS, Hybrid Intelligence Departement, Sankt Augustin, Germany\\
}
\IEEEauthorblockA{
  \IEEEauthorrefmark{2}%
PricewaterhouseCoopers GmbH, Actuarial Risk Modelling, Munich, Germany
}
\IEEEauthorblockA{
  \IEEEauthorrefmark{3}%
  loong.kuan.lee@iais.fraunhofer.de\\
}
}

\maketitle
\IEEEpeerreviewmaketitle


\begin{abstract}
The rising energy production costs and the increasing reliance on volatile renewable sources have driven the need for more efficient power system redispatch strategies. In this work, we re-interpret the redispatch problem as a multi-objective combinatorial optimization task within the Quadratic Unconstrained Binary Optimization (QUBO) framework, suitable for adiabatic quantum computing. Our contributions include a novel normalized unbalanced penalty method that integrates inequality constraints via a quadratic Taylor expansion and an $\alpha$-Expansion algorithm that allows us to address large-scale redispatch instances and to integrate temporal adjacent state switching constraints directly into the algorithm. Our experiments are conducted on open data of the German power system. Our results, obtained via numerical simulation and from an actual D-Wave Advantage quantum annealer, validate the viability of our formulation and demonstrate that our algorithm scales to large problem instances. 
\end{abstract}


\section{Introduction}
\label{sec:introduction}
The intelligent control of power systems gained increased importance with the rising 
cost of energy production and the retirement of fossil energy sources. As traditional 
fossil fuel-based power plants are phased out due to environmental regulations and 
sustainability goals, the power system is increasingly relying on renewable energy sources 
such as wind and solar. While these renewable sources are cleaner, they are also more volatile 
and less predictable, which complicates the task of balancing supply and demand. 

Simultaneously, the cost of energy production is rising due to several
factors, including higher fuel prices, increased operational costs, and
the investment required to integrate renewable energy sources into the
grid. This scenario necessitates a more dynamic and cost-effective
approach to energy dispatch. Efficient redispatch strategies are crucial
to minimize production costs while ensuring that the power system
remains reliable and stable. This involves optimizing the output of
available generators, considering the volatile nature of renewable
energy, and managing the constraints of the power system.  By addressing
these challenges, power system redispatch helps in maintaining a balance
between economic efficiency and the transition towards a more
sustainable energy future.

Mathematical formulations of this scenario turn out to be hard combinatorial optimization problems. 
These problems involve numerous variables and constraints, making them difficult to solve. 
Assuming P$\not =$NP, we do not expect that an efficient algorithm exists for solving 
these problems optimally in polynomial time. As a result, for large-scale problem instances, no 
efficient solvers currently exist that can provide optimality guarantees within a practical timeframe. 
This computational challenge underscores the need for advanced optimization techniques and heuristics 
to provide solutions that are both feasible and effective in real-world scenarios. 

The best currently known quantum speed-up for unstructured search (combinatorial optimization) 
is quadratic---via Grover's search~\cite{grover96search,grover2025amplify}. 
In practice, the speed-up is hard to obtain on the current generation of universal 
quantum computers. However, 
quantum annealing~(QA)~\cite{farhi2000quantumcomputationadiabaticevolution} evolved as a 
practical alternative to address high-dimensional 
combinatorial optimization problems.

\subsection{Practical Quantum Computing} \label{sec:qc}
Let us quickly introduce the basic notion of what can be considered as quantum computation~\cite{Nielsen/Chuang/2016a}. 
Today, practical quantum computing (QC) consists of two dominant paradigms: Adiabatic QC (AQC) and gate-based QC. In both scenarios, a quantum state $\ket{\psi}$ for a system with $n$ qubits is a $2^n$ dimensional complex 
vector. In the gate-based (universal) framework, a quantum computation is defined as a matrix multiplication $\ket{\vec{\psi}_{\text{out}}}=C\ket{\vec{\psi}_{\text{in}}}$, where the $C$ is a $(2^n \times 2^n)$-dimensional unitary matrix (the circuit), given via a 
series of inner and outer products of low-dimensional unitary matrices (the gates). The circuit is derived manually or via heuristic search algorithms~\cite{Franken2022}. In AQC---the
framework that we consider in the paper at hand---the result of computation is defined to be the eigenvector $\ket{\phi_{\min}}$ that corresponds to the smallest eigenvalue of some $(2^n \times 2^n)$-dimensional Hermitian matrix $\mat{H}$. In 
practical AQC, $\mat{H}$ is further restricted to be a real diagonal matrix whose entries can be 
written as $\mat{H}_{i,i}=\mat{H}(\mat{Q})_{i,i}=\boldsymbol{\vec{x}^{i{\trans}} \mat{Q} \vec{x}^i}$ where $\vec{x}^i=\operatorname{binary}(i)\in\{0,1\}^n$ is some (arbitrary but 
fixed) $n$-bit binary expansion of the unsigned integer $i$. Here, $\mat{Q} \in \mathbb{R}^{n\times n}$ is the so-called {\QUBO} (Quadratic Unconstrained Binary Optimization) 
matrix. By construction, computing $\ket{\vec{\psi}_{\text{out}}}$ is equivalent to solving 
\begin{equation}\label{qubo-objective}
	\min_{\vec{x}\in\{0,1\}^n} \vec{x}^{\trans} \mat{Q} \vec{x}\;.
\end{equation}Adiabatic quantum algorithms rely on this construction by encoding (sub-)problems as {\QUBO} matrices. 

In both paradigms, the output vector is $2^n$-dimensional and can thus not be read-out efficiently for non-small $n$. Instead, the output of a practical quantum computation is a random integer $i$ between 1 and $2^n$, which is drawn from the probability mass function $\operatorname{Prob}(i)=|\braket{i | \vec{\psi}_{\text{out}}}|^2=|\ket{\vec{\psi}_{\text{out}}}_i|^2$. This sampling step is also known as collapsing the quantum state $\ket{\vec{\psi}_{\text{out}}}$ to a classical binary state $\operatorname{binary}(i)$.

AQC has been applied to numerous hard combinatorial optimization problems~\cite{lucas2014}, ranging over satisfiability~\cite{kochenberger2005}, routing problems~\cite{neukart2017} to machine learning~\cite{bauckhage2019}.

Therefore, our contributions will involve and developing a formulation
and an approach for finding solutions to the power system redispatch
problem using AQC. More specifically, our contributions can be
summarized as follows:
\begin{itemize}
\item We provide a multi-objective formalization of the redispatch
  problem in the framework of {\QUBO}.
\item We devise a novel normalized version of unbalanced penalty for
  integrating inequality constraints into any {\QUBO}, by leveraging
  restrictions known of the problem.
\item We formulate a novel $\alpha$-Expansion algorithm for optimizing
  very large redispatch instances. In particular, we integrate the
  temporal adjacent state switching constraint that arises from
  the problem formulation into our iterative algorithm procedure and not
  into the objective
  function. 
\item We conduct numerical experiments which confirm our theoretical
  considerations, and show the applicability of our approach
  in~\cref{sec:multi-obj} for applying scalarization weights from small
  problem sizes to larger problem sizes.
\item Finally, we show that our proposed approach allows us to
  address problem sizes that are otherwise infeasible on current day
  quantum annealers by applying our approach on a D-Wave
  Advantage System 4.1 quantum annealer.
\end{itemize}

\section{Related Work}
In recent years, there has been a growing interest in addressing
challenges related to economic dispatch and power system
optimization~\cite{ajagekar2019energy}.  The comprehensive survey
in~\cite{ciornei2013recent} summarized the developments in economic
dispatch over the past two decades and categorizing the research based
on market structures and variable generation sources.

In the realm of combinatorial optimization, Li et
al.~\cite{li2003generation} addressed generation scheduling with thermal
stress constraints using Lagrangian relaxation and simulated annealing,
examining the economic costs of frequent ramping and its effects on
turbine lifespan.  Lin et al.~\cite{lin2001nonconvex} tackled the
nonconvex economic dispatch problem with an algorithm combining
evolutionary programming, tabu search, and quadratic programming,
demonstrating improved performance over prior evolutionary methods.
Tanjo et al.~\cite{tanjo2016graph} developed a graph partitioning
algorithm for decentralized electricity management of renewable energy,
minimizing costs associated with resilient microgrid construction by
optimizing surplus electricity transfer between microgrids.  Knueven et
al.~\cite{knueven2020mixed} presented a survey of mixed integer
programming formulations for the unit commitment problem in power grid
operations, while also introducing novel formulations to improve
performance.

In the field of quantum computation, power system related approaches
include a novel method to optimize electricity surplus in transmission
networks using QA~\cite{colucci2023power}. Specifically the authors
demonstrated that quantum-classical hybrid solvers can outperform
classical methods in optimizing electricity surplus in transmission
networks. Similarly, Ajagekar et al.~\cite{ajagekar2022hybrid} presented
QC-based strategies for large-scale scheduling in manufacturing,
tackling mixed-integer programs with efficient classical-quantum hybrid
techniques. Stollenwerk et al.~\cite{stollenwerk2020toward} formulated
planning problems in terms of optimization problems over proper
colorings of chordal graphs. They introduced an efficient Quantum
Alternating Operator Ansatz (QAOA)~\cite{farhi2014} contruction to
achieve resource scaling as low-degree polynomials of input
parameters. Finally, Nikmehr et al.~\cite{nikmehr2022quantum} formulated
a quantum unit commitment model and a quantum distributed approach for
large-scale discrete optimization, showcasing the potential of quantum
computing in complex power systems.



These works collectively provide a foundation for addressing economic
dispatch and power system optimization challenges. For general
information on power generation, control, dispatch, planning, and
scheduling we refer to~\cite{wood1996power}
and~\cite{pinedo2005planning}.

\section{Background} \label{sec:background}

We start off with the notation used throughout this paper in
\cref{sec:notation} and then move to describing the redispatch problem
in \cref{sec:redispatch}.

\subsection{Notation}\label{sec:notation}

We denote matrices with bold capital letters (e.g. $\mat{A}$) and
vectors with bold lowercase letters (e.g. $\vec{a}$). Furthermore, for a
vector $\vec{a}\in\mathbb{R}^{k_{1}\cdot k_{2} \cdot\ldots\cdot k_{n}}$,
we allow the indexing of the vector via the notation
$\vec{a}_{i_{1},i_{2},\ldots,i_{n}}$ in order to get the
$(i_{1}\cdot (k_{2}\times\ldots\times k_{n}) + i_{2}\cdot
(k_{3}\times\ldots\times k_{n}) + \ldots +i_{n})$-th element of
$\vec{a}$. On the other hand we define $\mat{A}_{i,\cdot}$ and
$\mat{A}_{\cdot,j}$ to be the $i$-th row and $j$-th column of the matrix
$\mat{A}$ respectively. We also define $\text{abs}_{\odot}(\bx)$ as the
element-wise absolute value of some vector $\bx$.

Furthermore, we use the following standard terms of linear algebra. Let
$\diag\left(\vec{a}\right)$ be the $n\times n$ diagonal matrix with
$\vec{a}\in\mathbb R^n$ as its diagonal. Additionally, we define
$\mat{I}_n$ as the $n$-dimensional identity matrix, $\ones{n}$ to be the
$n$-dimensional vector consisting only of $1$s. Furthermore, we will use
$\iverson{\cdot}$ to denote the Iverson bracket where for some statement
$P$, $\iverson{P}=1$ if $P$ is true. Lastly, we use $\ups{n}$ to refer
to the $n\times n$ upper shift matrix defined as
$[\ups{n}]_{ij} = \iverson{i=j+1}$.

\subsection{Redispatch} \label{sec:redispatch}

Consider $n$ controllable resources, each with $k$ possible states,
across $T$ discrete timepoints. Let $\bZ$ be a $T\times n$ matrix over
the index set $\indices{k}=\{1,\dots,k\}$ represent a specific
configuration of all controllable resources over time.  The power
produced by resource $a$ at time $t$ is
\begin{gather}
  \label{eq:8}
  [\g{\bP}(\g{\bZ})]_{t,a}=
  \sum_{i=1}^k \iverson{\g{\bZ}_{t,a} = i}\cdot\g{\bp}_{a,i}\;,
\end{gather}
where $\bp_{a,i}\geq 0$ is the power produced by operating resource
$a\in\indices{n}$ in state $i\in\indices{k}$ such that
$\forall i \in \indices{k-1} : \bp_{a,i} < \bp_{a,i+1}$.

At each timepoint $t\in\indices{T}$, we are given a forecasted amount of
power $\btau_{t}$ to be produced in total. In order to maintain the
stability of the power system, we need to ensure that the total power
produced by all $n$ controllable resources matches the given target
$\btau$ as close as possible,
\begin{equation}
  \label{eq:constraint-power}
  \forall t \in\indices{T} : 
  \sum_{a=1}^{n}[\g{\bP}(\g{\bZ})]_{t,a}
  \approx \btau_{t}
  \iff
  \g{\bP}(\g{\bZ}) \ones{n}
  \approx \btau\;.
\end{equation}
Slight deviations from $\btau$ can be absorbed in various ways such as
by slight changes in the frequency of the power system.

Furthermore, each controllable resource has a fixed ramp rate,
representing the maximum upward or downward change in energy production
per timepoint. To model this, we assume that for any pair of adjacent
timepoints $t, t+1 \in \indices{T}$, a controllable resource
can transition by at most one state, 
\begin{equation}
  \label{eq:constraint-ramp}
  \forall t \in\indices{T-1} : 
  \text{abs}_{\odot}([\bZ]_{t,\cdot} - [\bZ]_{t+1,\cdot})
  \leq \ones{n}\;,
\end{equation}
where $\text{abs}_{\odot}(\bx)$ is the element-wise absolute value of a
vector.

The redispatch problem then seeks to minimize the total cost of an
energy network configuration $\bZ$, subject to the constraints in
\cref{eq:constraint-power,eq:constraint-ramp}. The cost associated with
a configuration $\bZ$ stems from three sources. The first source is the
production cost of running the $a$-th controllable resource at timepoint
$t$ in its $i$-th state; this incurs a non-negative cost
$\bc_{t,a,i}>0$. The total production cost for a $\bZ$ is then:
\begin{equation}
  \label{eq:cost}
  C(\g{\bZ})=\sum_{t=1}^T\Biggl(
  \sum_{a=1}^n \sum_{i=1}^{k} \g{\bc}_{t,a,i} \cdot
  \delta\Bigl([\bZ]_{t,a},i\Bigr)
  \Biggr)\;.
\end{equation}

Furthermore, sending a redispatch order to a controllable resource
incurs a cost associated with adjusting the production levels. This cost
is generally proportional to the change in energy production when
transitioning the $a$-th controllable resource from state $i$ to $i'$---
$\bs_{a}(i, i') = \gamma \bigl|\g{\bp}_{a,i} - \g{\bp}_{a,i'}\bigr|$,
where $\gamma$ is some arbitrary constant. Therefore, the cost of the
redispatch orders in configuration $\bZ$---which we will call the
\textit{switching cost} of $\bZ$---is:
\begin{gather}
  \label{eq:cost-switching}
  W(\g{\bZ})
  = \sum_{t=1}^{T-1}\sum_{a=1}^{n}
  \bs_{a}(\g{\bZ}_{t,a}, \g{\bZ}_{t+1,a})\;.
\end{gather}

The final source of cost for a configuration $\bZ$ over the controllable
resources of a power system comes from overloaded transmission lines in
the system. Overloading transmission lines accelerate wear, leading to
premature replacements being required.  To encode this objective, assume
we are given the sensitivity matrix $\bS$ with dimensions $n\times L$,
where $L$ is the number of transmission lines. $\bS$ quantifies the
power flow from each controllable resource $a\in\indices{n}$ through
each transmission line $l\in\indices{L}$. Also assume we have a
non-negative $T\times L$ matrix $\bM$ representing the time-dependent
rated maximum load for each transmission line. This rated capacity can
change over time due to factors such as the external temperature
at different timepoints~\cite{karimi2018}. The number of overloaded
transmission lines can then be expressed as:
\begin{equation}
  \label{eq:cost-overload}
  O(\bZ) = \sum_{t=1}^{T}\sum_{l=1}^{L}
  \iverson{[\bP(\bZ)\bS]_{t,l} \geq [\bM]_{t,l}}\;.
\end{equation}

Therefore, the final multi-objective problem we wish to minimize is:
\begin{alignat}{2}
  \min
  & \quad
    C(\g{\bZ}), W(\g{\bZ}), O(\bZ) && \label{eq:obj}\\
  \text{subject to}
  & \quad
    \g{\bP}(\g{\bZ}) \ones{n} \approx \btau&&
    \label{eq:con-power} \\
  &\quad
    \forall t\in\indices{T-1} :
  \text{abs}_{\odot}([\g{\bZ}]_{t,\cdot} - [\g{\bZ}]_{t+1,\cdot})\;. &&
      \label{eq:con-adjacent}
\end{alignat}
It should be mentioned that an additional implicit constraint here is
the plain integer constraint on the configuration variables $\bZ$.

\section{Problem Formulation}
\label{sec:naive}
To solve the problem defined in \cref{sec:redispatch} on AQC hardware, we
must first formulate it as a {\QUBO}. To this end, we rewrite the
problem using binary variables $\bx$ representing a vectorized one-hot
encoding of the matrix $\bZ$.
\begin{equation}
  \label{eq:one-hot-encoding}
  \bx_{t,a,i}
  = \delta( [\g{\bZ}]_{t,a}, i )
  =
  \begin{cases}
    1 & \text{if }[\bZ]_{t,a} = i\\
    0 & \text{otherwise}
  \end{cases}\;.
\end{equation}
Therefore, the power produced by the $a$-th controllable resource at
time $t$ becomes
\begin{equation}
  \label{eq:7}
  [\g{\bP}(\g{\bx})]_{t,a}=\sum_{i=1}^k
  \g{\bx}_{t,a,i}\cdot\g{\bp}_{a,i}\;.
\end{equation}
For the rest of this section, we re-formulate the objective function in
\cref{eq:obj} and constraints in \cref{eq:con-power,eq:con-adjacent}
using the binary vector $\bx$. We then demonstrate how this
reformulation can be expressed as a {\QUBO} problem.

\subsection{Objective Functions}
Formulating the production and switching costs in terms of $\bx$---and
thus expressing them as matrices within a {\QUBO} problem---is
relatively straightforward. These matrices can be directly constructed
from the linear and quadratic terms of their respective expressions with
respect to $\bx$.

\subsubsection{Production Cost}
Starting with the the production cost in \cref{eq:cost}, converting it
from an expression in terms of $\bZ$ to one in terms of $\bx$ gives us
\begin{gather}
  \label{eq:cost-bin}
  \begin{split}
    C(\bx)
    &= \sum_{t=1}^{T}\Biggl(
    \sum_{a=1}^{n}\sum_{i=1}^{k}\g{\bc}_{t,a,i} \g{\bx}_{tai}
      \Biggr)
    =\bx^{\trans} \diag(\g{\bc}) \bx
  \end{split}\;,
\end{gather}
and therefore the {\QUBO} matrix $\bQ_{C}=\diag(\bc)$.
\subsubsection{Switching Cost}
Formulating the switching cost as a {\QUBO} can be done similarly by
rewriting \cref{eq:cost-switching} in terms of the binary variables
$\bx$,
\begin{equation}
  \label{eq:cost-switching-bin}
  \begin{split}
    &W(\bx)\\
    &= \sum_{t=1}^{T-1}\sum_{a=1}^{n}
    \Biggl| \sum_{i=1}^{k}\g{\bp}_{t.a,i}\g{\bx}_{t,a,i} -
      \sum_{i=1}^{k}\g{\bp}_{t+1,a,i}\g{\bx}_{t+1,a,i} \Biggr|\\
    &=\sum_{t=1}^{T-1}\sum_{a=1}^{n}
      \sum_{i=1}^{k}\sum_{i'=1}^{k}
      \g{\bx}_{t,a,i} \g{\bx}_{t+1,a,i'}
      \bigl| \g{\bp}_{t,a,i} - \g{\bp}_{t+1,a,i'} \bigr|\;,
  \end{split}
\end{equation}
and therefore allowing for the direct mapping of the quadratic terms in
\cref{eq:cost-switching-bin} to the {\QUBO} matrix $\bQ_{W}$,
\begin{gather}
  \label{eq:26}
  \bigl[\bQ_{W}\bigr]_{(t,a,i),(t+1,a,i')}
  = \bigl| \g{\bp}_{t,a,i} - \g{\bp}_{t+1,a,i'} \bigr|\;.
\end{gather}

\subsubsection{Overload Cost}\label{sec:cost-overload}
Unfortunately, formulating the objective function $O(\bZ)$ in
\cref{eq:cost-overload} as a {\QUBO} matrix is not as
straightforward. As a first step, we can re-express $O(\bZ)$ in terms of
$\bx$:
\begin{align}
  \label{eq:cost-overload-bin}
  O(\bx)
  &= \sum_{t=1}^{T}\sum_{l=1}^{L}
           \iverson{[\bP(\bx)\bS]_{t,l} \geq [\bM]_{t,l}}\\
  &= \sum_{t=1}^{T}\sum_{l=1}^{L}
    \iverson{\sum_{a=1}^{n} [\bS]_{a,l}
    \sum_{i=1}^{k}\bx_{t,a,i}\cdot\bp_{a,i}
    \geq [\bM]_{t,l}}\;.\nonumber
\end{align}
In order to re-formulate $O(\bx)$ into a {\QUBO} matrix, we first need
to find some linear or quadratic approximation---with respect to
$\bx$---of the inequalities in the Iverson brackets of~\cref{eq:cost-overload-bin} for all $t\in\indices{T}$ and
$l\in\indices{L}$,
\begin{gather}
  \label{eq:cost-overload-ineq}
  \sum_{a=1}^{n}[\bS]_{a,l}\sum_{i=1}^{k}\bx_{t,a,i}\bp_{a,i}
  \geq [\bM]_{t,l} \iff h_{t,l}(\bx) \leq 0\;,
  \intertext{where}
  \label{eq:cost-overload-h}
  h_{t,l}(\bx) =
  [\bM]_{t,l} -
  \sum_{a=1}^{n}[\bS]_{a,l}\sum_{i=1}^{k}\bx_{t,a,i}\cdot\bp_{a,i}\;.
\end{gather}
A good approximation of $\iverson{h_{t,l}(\bx)\leq 0}$ would then need
to be a function on $h_{t,l}(\bx)$ that returns large values when
$h_{t,l}(\bx)\leq 0$ and small values otherwise.
\begin{figure}
  \centering
  \includegraphics[width=0.8\columnwidth]{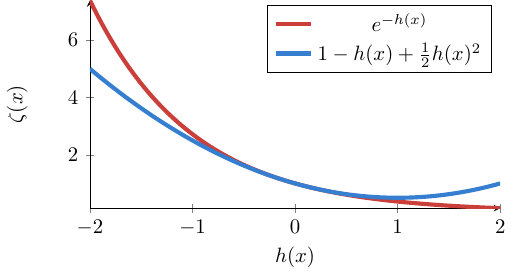}
  \caption{Negative exponential and its Taylor approximation to the
    second order.\label{fig:taylor-approx}}
\end{figure}

One approach would be to use unbalanced
penalization~\cite{montanez-barrera2023a} and approximate
$\iverson{h_{t,l}(\bx)\leq 0}$ with the negative exponential function
$\iverson{h_{t,l}(\bx)\leq 0}\approx e^{-h_{t,l}(\bx)}$. To formulate
the negative exponential as a {\QUBO} matrix, we will further
approximate it by the second order Taylor expansion around
$h_{t,l}(\bx)=0$,
\begin{align}
  \label{eq:up-ori}
     e^{-h_{t,l}(\bx)}
      \approx \zeta(\bx)
    &=1 - h_{t,l}(\bx) + \frac{1}{2} h_{t,l}(\bx)^{2}\;,
\end{align}
resulting in a quadratic approximation of
$\iverson{h_{t,l}(\bx)\leq 0}$. The difference between $e^{-x}$ and its
quadratic approximation is illustrated in \cref{fig:taylor-approx}.

However, the second-order Taylor approximation of $e^{-h_{t,l}(\bx)}$
around $h_{t,l}(\bx)=0$ is accurate only within the range
$-1\leq h_{t,l}(\bx)\leq 1$. To minimize approximation errors, we
propose normalizing $h_{t,l}(\bx)$. Specifically, we only need to
normalize $h_{t,l}(\bx)$ such that $h_{t,l}(\bx) \leq 1$; this is
because we don't require an accurate approximation of
$e^{-h_{t,l}(\bx)}$ for negative values of $h_{t,l}(\bx)$, as we only
require $\zeta(\bx) > e^{0}=1$ in that region.

Therefore, to guarantee that $h_{t,l}(\bx)\leq 1$, we need an upper
bound on it. Fortunately, we know that the negative term in
\cref{eq:cost-overload-h}---which represents the load on the
transmission line $l$ at time $t$---is by definition minimized if every
controllable resource is at its lowest power production
state. Therefore, $h_{t,l}(\bx)$ is maximized at this lowest power
production configuration, leading the to upper bound
\begin{equation}
  \label{eq:cost-overload-h-max}
  \fnorm{h_{t,l}} \defeq \max_{\bx} h_{t,l}(\bx) =
  [\maxload]_{t,l} - \sum_{a=1}^{n}[\bS]_{a,l}\cdot\bp_{a,1}\;,
\end{equation}
for all $t \in \indices{T}, l \in \indices{L}$.

Therefore the final approximation of $\iverson{h_{t,l}(\bx) \leq 0}$ can
be expressed as
\begin{equation}
  \label{eq:3}
  \iverson{h_{t,l}(\bx) \leq 0}
  \approx
  1 - \frac{h_{t,l}(\bx)}{\fnorm{h_{t,l}}}
  + \frac{1}{2}\cdot \frac{h_{t,l}(\bx)^{2}}{\fnorm{h_{t,l}}^{2}}\;.
\end{equation}
Upon substitution into \cref{eq:cost-overload-bin}, this yields an
approximation of the objective function $O(\bx)$ that can be formulated
as a {\QUBO} objective function,
\begin{align}
  O(\bx)
  \approx & \sum_{t=1}^{T}\sum_{a=1}^{n}\sum_{i=1}^{k}
            \bx_{t,a,i} \bv_{t,a,i}\label{eq:cost-overload-qubo}\\
          &+ \sum_{t=1}^{T}\sum_{a=1}^{n}\sum_{i=1}^{k}
            \sum_{b=1}^{n}\sum_{j=1}^{k}
            \bx_{t,a,i}\bx_{t,b,j} [\bW]_{(t,a,i),(t,b,j)}\;,\nonumber
\end{align}
where
\begin{gather}
  \label{eq:4}
  \bv_{t,a,i} = \bp_{a,i} \sum_{l=1}^{L} [\bS]_{a,l}
  \frac{\fnorm{h_{t,l}} -[\maxload]_{t,l}}{\fnorm{h_{t,l}}^{2}}\\
  [\bW]_{(t,a,i),(t,b,j)}
  = \bp_{a,i}\cdot\bp_{b,j}
  \sum_{l=1}^{L}
  \frac{[\bS]_{a,l}[\bS]_{b,l}}{2\fnorm{h_{t,l}}^{2}}\;.
\end{gather}
See \cref{fig:cost-overload-qubo} in the Appendix for a full derivation
of \cref{eq:cost-overload-qubo}. We can then formulate $O(\bx)$ as the
{\QUBO} matrix,
\begin{gather}
  \label{eq:1}
  \begin{aligned}
    &[\bQ_{O}]_{(t,a,i), (t,b,j)}\\
    &=
    \begin{cases}
      [\bW]_{(t,a,i),(t,b,j)} +\bv_{t,a,i} & \text{if } a = b, i=j\\
      [\bW]_{(t,a,i),(t,b,j)} & \text{otherwise}
    \end{cases}\;,
  \end{aligned}
\end{gather}
for all $t\in\indices{T}$, $a,b\in\indices{n}$, and $i,j\in\indices{k}$.

We now have a formulation of the objective functions for the
multi-objective {\QUBO} problem we wish to solve,
\begin{equation}
  \label{eq:init-qubo}
  \argmin_{\bx}\: \mu_{C}\bQ_{C}(\bx)
  + \mu_{S}\bQ_{S}(\bx)
  + \mu_{O}\bQ_{O}(\bx)\;.
\end{equation}
where $\mu_{C}$, $\mu_{S}$, and $\mu_{O}$ are weights given to the
objectives when adding them
together---i.e. scalarizing~\cite{ayodele2023} them---into a single
objective function.  However, we still need to formulate the constraints
of our problem in \cref{eq:con-power,eq:con-adjacent} as {\QUBO}
matrices.

\subsection{Soft Constraint: Energy Production Target}
Recall that when finding a configuration $\bx$ that minimizes the
overall costs within the power system, it is also crucial to ensure that
the resulting power generated by the configuration closely matches the
forecasted power target at each timepoint $\btau$. So for all
$t\in\indices{T}$,  we require:
\begin{align}
  \label{eq:constraint-power-bin}
  \sum_{a=1}^{n}[\g{\bP}(\bx)]_{t,a} \approx \btau_{t}
  \implies
  \bP(\bx)\ones{n} \approx \btau\;.
\end{align}

We refer to this constraint as a \textit{soft} constraint because it
does not strictly enforce equality between the total power generated and
the forecasted power target. Instead, the objective is to minimize the
``distance'' between the total power produced at each timepoint and the
given target; this minimization problem can be formulated as the {\QUBO}
problem,
\begin{align}
  \argmin_{\bx}\|
  \bP(&\bx)\ones{n}- \btau\|_{2}^{2}\nonumber\\
    =\argmin_{\bx}&\sum_{t=1}^{T} \Biggl(
    \sum_{a=1}^{n}[\bP(\bx)]_{t,a} - \btau_{t}\Biggr)^{2}
    \nonumber\\
  =\argmin_{\bx}
  &\sum_{t=1}^{T}
    \sum_{a=1}^{n}\sum_{b=1}^{n} \sum_{i=1}^{k}\sum_{j=1}^{k}
    \bx_{t,a,i}\cdot\bx_{t,b,j\cdot}\bp_{a,i}\cdot\bp_{b,j}
    \nonumber\\
  &-2\sum_{t=1}^{T}\sum_{a=1}^{n}\sum_{i=1}^{k}
    \bx_{t,a,i}\cdot\bp_{a,i}\cdot\btau_{t}
    \nonumber\\
  =\argmin_{\bx} &\:\bigl(\bx^{\trans} \bQ_{P}\: \bx\bigr)\;,
                \label{eq:power-qubo-eq}
\end{align}
where $\|\cdot\|_{2}^{2}$ is the squared $L^{2}$-norm of a vector and,
\begin{equation}
  \label{eq:power-qubo-mat}
  \begin{aligned}
    &[\bQ_{P}]_{(t,a,i),(t,b,j)}\\
    &=
      \begin{cases}
        \bp_{a,i} \cdot \bp_{b,j} -2 \cdot \bp_{a,i} \cdot \btau_{t}
        & \text{if } a=b, i=j\\
        \bp_{a,i} \cdot \bp_{b,j} &\text{otherwise}
      \end{cases}\;,
  \end{aligned}
\end{equation}
for all $t\in\indices{T}$, $a,b\in\indices{n}$, and $i,j\in\indices{k}$.

\subsection{Hard Constraints}\label{sec:qubo-hard-cons}
Finally, there are a couple of constraints that either cannot be, or we
do not wish to be, violated at all cost; therefore we refer to these set
of constraints as \textit{hard} constraints.

\subsubsection{One-Hot Constraint}
The first hard constraint we need to consider comes from representing a
configuration $\bZ$ as a one-hot encoding of binary variables
$\bx$. Recall that $\bZ$ is a $T\times n$ matrix of integers in
$\indices{k}$. Under a one-hot encoding, each element in $\bZ$ is
encoded as a vector of $k$ binary variables, each representing an
integer in $\indices{k}$ when the respective binary variable is set to
$1$. Therefore, $\bx$ contains $T\times n$ of these one-hot encodings,
resulting in a vector of length $T\times n \times k$ with the constraint
that for all $t\in\indices{T}, a\in\indices{n}$,
\begin{align}
  \label{eq:one-hot-constraint}
  \sum_{i=1}^{k} \g{\bx}_{t,a,i} = 1
  &\iff \bigl(\ident{T\times n} \otimes \ones{k}^{\trans}\bigr) \bx
    =\ones{T\times n}\\
  &\implies
    \Bigl\lVert\bigl(\ident{T\times n} \otimes \ones{k}^{\trans}\bigr)
    \bx - \vec{1}_{T\times n}\Bigr\rVert_{2}^{2}=0\;,
      \label{eq:con-one-hot}
\end{align}
where $\otimes$ is the Kronecker product. This constraint can then be
expressed as the minimization problem
\begin{align}
  &\argmin_{\bx}\: \Bigl\lVert\bigl(
  \ident{T\times n} \otimes \ones{k}^{\trans}\bigr) \bx -
    \vec{1}_{T\times n}\Bigr\rVert_{2}^{2}\nonumber\\
  &=\argmin_{\bx}\: \sum_{t=1}^{T}\sum_{a=1}^{n} \Biggl(
    \sum_{i=1}^{k}\bx_{t,a,i} - 1 \Biggr)^{2}\nonumber\\
  &=\argmin_{\bx}\: \sum_{t=1}^{T}\sum_{a=1}^{n} \Biggl(
    \sum_{i,j=1}^{k}\bx_{t,a,i}\bx_{t,a,j}
    - 2\sum_{i=1}^{k}\bx_{t,a,i}\Biggr)\nonumber\\
  &=\argmin_{\bx} \: \bx^{\trans} \bQ_{H} \bx\;,
\end{align}
where
\begin{equation}
  \label{eq:one-hot-qubo-mat}
  [\bQ_{H}]_{(t,a,i),(t,a,j)} =
  \begin{cases}
    -1 & \text{if } i = j\\
    1 & \text{otherwise}
  \end{cases}\;,
\end{equation}
for all $t\in\indices{T}$, $a\in\indices{n}$, and $i,j\in\indices{k}$.
\subsubsection{Adjacent State Switching Constraint}
\label{sec:constr-adj}
Another constraint that must not be violated is the requirement that all
controllable resources can only switch to adjacent states between
consecutive timepoints. This adjacency constraint reflects a physical
limitation on the rate at which the controllable resources can ramp its
power production up or down; therefore violations would lead to
changes in power production that cannot be realized.

Recall from \cref{eq:con-adjacent} that we constrained any configuration
$\bZ$ such that $|\g{\bZ}_{t,a} - \g{\bZ}_{t+1,a}| \leq 1$ for all
$a\in\indices{n}$ and $t \in\indices{T-1}$.  Let $\by_{t}$ and
$\by_{t+1}$ be the one-hot binary encoding of integers $\g{\bZ}_{t,a}$
and $\g{\bZ}_{t+1,a}$ respectively. Then the objective function that
satisfies this constraint---i.e. that is at its minimum for solutions that
satisfy the constraint---is:
\begin{gather}
  \label{eq:con-adj}
  \min_{\by_{t},\by_{t+1}} \by_{t}^{\trans} \adjswitch \, \by_{t+1}\;,
\end{gather}
where $[\adjswitch]_{i,i'} = \iverson{|i - i'| > 1}$.  Therefore, the
final {\QUBO} problem to maintain the constraint
$|\g{\bZ}_{t,a} - \g{\bZ}_{t+1,a}| \leq 1$ over the entire stacked
one-hot encoding $\bx$ can be constructed with the Kronecker product:
\begin{equation}
  \label{eq:9}
  \argmin_{\bx} \: \bx^{\trans}(\ups{T} \otimes \ones{n} \otimes \adjswitch)\bx
  =\argmin_{\bx} \: \bx^{\trans}\bQ_{A}\bx \;,
\end{equation}
where $\bQ_{A}=\ups{T} \otimes \ones{n} \otimes \adjswitch$ and $\ups{T}$
is the upper shift matrix. This is equivalent to duplicating the
constraint in \cref{eq:con-adj} to be applied on every pair of $\by_{t}$
and $\by_{t+1}$ in $\bx$.

\subsection{Final Problem Formulation}
Therefore, the final {\QUBO} of our redispatch problem is:
\begin{equation}
  \label{eq:fin-obj}
  \begin{aligned}
    \argmin_{\bx}\quad
    &\bx^{\trans} \Bigl(
      \mu_{C}\cdot\bQ_{C} + \mu_{S}\cdot\bQ_{S} + \mu_{O}\cdot\bQ_{O}
      \Bigr)\, \bx\\
    &+\bx^{\trans} \Bigl(
       \lambda_{H} \cdot\mat{Q}_{H} +
      \lambda_{A}\cdot\mat{Q}_{A}\Bigr)\, \bx\\
    & + \lambda_{P} \cdot \bx^{\trans}\mat{Q}_{P}\,\bx\\
    =\argmin_{\bx}& \:\:\bx^{\trans}\bQ \bx + \bx^{\trans}\hat{\bQ} \bx
                    + \lambda_{P} \cdot \bx^{\trans}\mat{Q}_{P}\,\bx\;,
  \end{aligned}
\end{equation}
where $\lambda_{P}$, $\lambda_{H}$, and $\lambda_{A}$ are the Lagrangian
multipliers for our constraints.

While we have shown how each part of the problem can be encoded as a
{\QUBO} problem, it turns out that the resulting matrices $\mat{Q}$,
$\hat{\bQ}$, and $\mat{Q}_{P}$ can become rather large. In realistic
setups with 5 states per controllable resource, the number of binary
variables in our problem is in the order of $T\times 1690$, where $T$ is
the number of timepoints considered in the redispatch problem.
Additionally, state-of-the-art quantum annealers have limits on the
maximum size of the problem it can solve. We address these issues in the
following section.

\section{Decomposing QUBO Problem via
  $\alpha$-expansion} \label{sec:alpha}

The {\QUBO} problems from the previous section can become quite large
for realistic setups.  Therefore, we need to use decomposers in order to
split our problem into a set of smaller instances whose solutions can
then be combined to a global solution of the original problem. 
Various approaches for decomposing {\QUBO}s are known and a full review
of this topic is out of the scope of this work.  We instead focus on an
algorithmic approach called $\alpha$-expansion.

The $\alpha$-expansion algorithm iteratively refines an initial valid
solution $\bx_0$ by selectively applying permutations that exchange the
positions of $0$s and $1$s in the current solution. Formally, at each
iteration, a subset of all possible permutation matrices
$\{\mat{C}_{1},\ldots,\mat{C}_{c}\} \subseteq \mathbb{C}$ are applied to
the current solution $\hat{\bx}$.  This set $\mathbb{C}$ contains only
permutation matrices that, when applied to $\hat{\bx}$, yield another
valid solution; consequently $\mathbb{C}$ depends on both the problem
being solved and the current solution $\hat{\bx}$.

Therefore, instead of directly optimizing over the potentially large
number of decision variables $\bx$, we iteratively formulate and solve a
series of {\QUBO} subproblems. Each subproblem optimizes over a choice
of $c$ permutation matrices that---when applied to the current solution
$\hat{\bx}$---minimizes the original objective
function~\cite{benkner2021}, $\min_{\bm{\alpha}\in\{0,1\}^{c}}\bm{\alpha}^{\trans} {\mat{B}}\bm{\alpha}$ with 
\begin{gather}
  \label{eq:2}
  \begin{aligned}
    \bigl[{\mat{B}}\bigr]_{i,j}
    &=\begin{cases}
      E({\vec{c}}_{i}^{*},{\vec{c}}_{j}^{*})
      & \text{if } i\neq j\\
      E({\vec{c}}_{i}^{*},{\vec{c}}_{j}^{*}) +
      E({\vec{c}}_{i}^{*},{\bx}) +
      E({\bx},{\vec{c}}_{j}^{*})
      & \text{otherwise}
    \end{cases}\;,
  \end{aligned}\nonumber
\end{gather}
where 
$E(\vec{v},\vec{u}) \defeq \vec{v}^{\trans} \vec{Q} \vec{u}$, 
$\vec{c}_{i}^{*} \defeq (\mat{C}_{i} - \ident{T\times m\times k})\bx$ and $\mat{C}_{i}\in\mathbb{C}$.

The question now is what should the set of ``valid'' permutation
matrices $\mathbb{C}$ be? Previous literature has shown that the
one-hot constraint can be obeyed by limiting $\mathbb{C}$ to cycles
that swap the values between two binary variables that are of the same
one-hot encoding~\cite{benkner2021,gerlach2023}. In our problem, these
cycles represent changing the state of controllable resource $a$ at
timepoint $t$ from state $i$ to $i'$:
\begin{gather}
  \label{eq:12}
  \g{\cyclefunc}_{t,b,i}(i') \defeq
  \bigoplus_{(t',a')=(1,1)}^{(T,n)}
  \begin{cases}
    \swapmat_{k}(i,i') & t=t', a=a'\\
    \ident{k} & \text{else}
  \end{cases}\;,
  \intertext{where the matrix $\swapmat_{k}(i,i')$ is defined as:}
  \bigl[\swapmat_{k}(i,i')\bigr]_{i,i'} =
  \bigl[\swapmat_{k}(i,i')\bigr]_{i',i} = 1\\
  \text{and} \quad \forall j \in \indices{k} \setminus \{i,i'\} :
  \bigl[\swapmat_{k}(i,i')\bigr]_{j,j} = 1\;.
\end{gather}
However, in addition to the plain one-hot constraint from
\cref{eq:con-one-hot}, our problem also has the adjacent state
constraint in \cref{eq:con-adj}. Therefore, finding a suitable
$\mathbb{C}$ is a bit more involved.

Specifically, assume the current solution is $\hat{\bx}$---which
corresponds to the configuration $\hat{\bZ}$---and we want to change the
state of the $a$-th controllable resource at timepoint $t$ from
$\hat{\bZ}_{t,a}$ to $i'\in\indices{k}$. We shall denote this desired
state change as $S=(t,j,i')\in\mathbb{S}$ where
\begin{gather}
  \label{eq:state-changes}
  \mathbb{S} = \Bigl\{
  (t,j,i) \mid \forall t\in\indices{T},a\in\indices{n},
  i\in\indices{k}\setminus \{\hat{\bZ}_{t,a}\}
  \Bigr\}\;.
\end{gather}
The cycle that will permute $\hat{\bx}$ to a solution with the desired
state $S$ is then just $\cyclefunc_{t,j,i'}(\hat{\bZ}_{t,j})$. However,
the resulting solution from this permutation is not guaranteed to be
valid. Therefore, given a target state change $(t, j, i')\in\mathbb{S}$,
\cref{alg:rectify-non-adj} will create a permutation matrix
$\mat{C}$---composed of potentially multiple cycles---such that the
cycle $\cyclefunc_{t,j,i'}(\hat{\bZ}_{t,j})$ is one of the cycles in
$\mat{C}$ and that the application of $\mat{C}$ on $\hat{\bx}$ results
in a valid solution.

\begin{algorithm}[t!]
  \SetAlgoNoEnd
  \SetAlgoVlined
  \caption{Rectify Non-Adjacent Swaps.}\label{alg:rectify-non-adj}
  \KwIn{Current Solution --- $\g{\bZ}$,
    New State --- ($t$, $j$, $i$)}
  \KwOut{Cycles to Rectify Non-Adjacent Swaps --- $\g{\mat{C}}$}
  $c \gets \g{\bZ}_{t,j}$\\
  $\g{\mat{C}} \gets F_{t,j,c}(i)$\\
  $l \gets t - 1$\\
  \While{$\bigl|c - \g{\bZ}_{l,j}\bigr| > 1$ and $l \geq 1$}{
      $c\gets \g{\bZ}_{l,j}$\\
      $c^{*}\gets c - \iverson{\g{\bZ}_{l,j}<c} +
      \iverson{\g{\bZ}_{l,j}\geq c}$\\
      $\g{\mat{C}} \gets \g{\cyclefunc}_{l,a,c}(c^{*})\g{\mat{C}}$\\
      $l \gets l - 1$\\
    }
  $c \gets \g{\bZ}_{t,j}$\\
  $r \gets t + 1$\\
  \While{$\bigl|c - \g{\bZ}^{*}_{r,j}\bigr| > 1$ and $r \leq T$}{
      $c\gets \g{\bZ}_{r,j}$\\
      $c^{*}\gets c - \iverson{\g{\bZ}_{r,j}<c} +
      \iverson{\g{\bZ}_{r,j}\geq c}$\\
      $\g{\mat{C}} \gets \g{\cyclefunc}_{r,a}(c, c^{*})\g{\mat{C}}$\\
      $r \gets r + 1$\\
    }
  \Return{$\g{\mat{C}}$}
\end{algorithm}

When selecting a subset of possible state changes,
$\{S_{1},\ldots,S_{c}\}\subset\mathbb{S}$, it is crucial to ensure that
the resulting $c$ permutation matrices from \cref{alg:rectify-non-adj}
are pairwise disjoint—that is, they operate on disjoint sets of decision
variables in $\bx$. To guarantee this, we enforce a separation of at
least $k$ timepoints between any two state changes considered at each
iteration; where as usual, $k$ represents the number of possible states
for each controllable resource. The complete approach to employing
$\alpha$-expansion for our redispatch problem is detailed in
\cref{alg:a-exp-full}. The convergence criteria used in this
paper---unless stated otherwise---is a lack of improvement in the
solution after five iterations. We will also use a subproblem size of
$c=128$ for the experiments in \cref{sec:experiments}.

\begin{algorithm}
  \SetAlgoNoEnd
  \SetAlgoVlined
  \caption{Redispatch $\alpha$-Expansion.}\label{alg:a-exp-full}
  \KwIn{$\hat{\bx}$---Initial Solution, $c$---subproblem size,
    \texttt{solver}---subproblem solver}
  \KwOut{$\bx$---Modified Solution}
  \Repeat{Some convergence criteria is met}{
      \Repeat{Every state change in $\mathbb{S}$ has been considered at
          least once
        }{
          Choose random subset of $n$ states in $\mathbb{S}$
          whose rectified cycles (\cref{alg:rectify-non-adj})
          $\{\g{\mat{C}}_{1},\ldots,\g{\mat{C}}_{c}\}$ are all
          disjoint.\\
          Create subproblem {\QUBO} ${\mat{B}}$ with cycles
          $ \{\g{\mat{C}}_{1},\ldots,\g{\mat{C}}_{c}\}$ and solution
          $\hat{\bx}$\\
          $\bm{\alpha}^{*}\gets
          \argmin_{\bm{\alpha}}\bm{\alpha}^{\trans} {\mat{B}}
          \bm{\alpha}$ using \texttt{solver}\\
          Apply cycles selected by $\bm{\alpha}^{*}$ on ${\hat{\bx}}$
        }
      }
      \Return{$\hat{\bx}$}
\end{algorithm}

As a result, the search space of $\alpha$-expansion is limited to
solutions that do not violate our hard constraints. Therefore, we can
omit the penalties from the hard constraints during optimization
resulting in the following multi-objective {\QUBO} problem:
\begin{equation}
  \label{eq:fin-obj-true}
  \argmin_{\bx} \:\:
  \bx^{\trans} \Bigl(
  \mu_{C}\bQ_{C} + \mu_{S}\bQ_{S} + \mu_{O}\bQ_{O}
  + \lambda_{P} \mat{Q}_{P}
  \Bigr)\, \bx\;.
\end{equation}
The remaining question then, is how we should determine the weights
assigned to our different objectives—$\mu_{C}$, $\mu_{S}$, and
$\mu_{O}$—and the Lagrangian multiplier $\lambda_{P}$ associated with
our soft constraint.

\section{Finding Multi-Objective Weights}\label{sec:multi-obj}
\begin{algorithm}[t!]
  \SetAlgoNoEnd
  \SetAlgoVlined
  \caption{Averaging Adaptive
    Method~\cite{ayodele2023}}\label{alg:pf}
  \KwIn{$\bQ_{O}$, $\bQ_{C}$, $\bQ_{S}$, $\lambda_{P}$, $\bQ_{P}$,
    \texttt{n\_weights}, \texttt{n\_runs}}
  \KwOut{$A$ --- A set of solutions and their weights}
  $\Lambda \gets \{(1,0,0), (0,1,0), (0,0,1)\}$\\
  $A\gets\emptyset, W\gets\{\}$\\
  \For{$i \in \{1,\ldots,\texttt{n\_weights}\}$} {
    \If{$i\leq m$}{
      $\bm{\mu} \gets \Lambda_{i}$
    }\Else{
      sort $W$ by each tuple's weights $\bm{\mu}$\\
      select two adjacent weight vectors $\bm{u}$ and $\bm{v}$ from $W$
      with the largest Euclidean distance in objective space\\
      $\bm{\mu}\gets ((\bm{u}_{1}+\bm{v}_{1})/2,
      (\bm{u}_{2}+\bm{v}_{2})/2, (\bm{u}_{3}+\bm{v}_{3})/2)$
    }
    $\bQ \gets \bm{\mu}_{1}\bQ_{O} + \bm{\mu}_{2}\bQ_{C} +
    \bm{\mu}_{3}\bQ_{S} + \lambda_{P}\bQ_{P}$\\
    $\bX \gets$solution set from running \cref{alg:a-exp-full} on $\bQ$
    \texttt{n\_runs} times.\\
    add to $A$ elements in the set $\{(\bx, \bm{\mu}) : \bx \in \bX\}$\\
    $\bx \gets $ solution in $\bX$ with smallest objective value\\
    $W_{i} \gets (\bm{\mu}, \bx, (\bx^{\trans}\bQ_{O}\,\bx,
    \bx^{\trans}\bQ_{C}\,\bx, \bx^{\trans}\bQ_{S}\,\bx))$\\
  }
  \Return{$A$}
\end{algorithm}

Although we do have a preference on which objectives we wish to
prioritize, specifically in the order of
\begin{enumerate}
\item the soft constraint on the power target $\bQ_{P}$,
\item the number of overloaded transmission lines $\bQ_{O}$
\item the production cost of the configuration $\bQ_{C}$, and
\item the switching cost of the configuration $\bQ_{S}$,
\end{enumerate}
it is often difficult to express these preferences as weights for a
multitude of reasons such as the objectives having different
scales~\cite{miettinen2008}. Therefore, we adopt an a posteriori
approach where, we first find a set of solutions and their
corresponding weights by solving multiple scalarizations of our
objectives using \cref{alg:pf}. The parameters $\lambda_{P}$ and
\texttt{n\_weights} used in \cref{alg:pf} are discussed further in
\cref{sec:experiments}.

We then choose a solution from this set $A$ based on our
preferences. Specifically, we first filter out any solution that results
in power production that deviates more than $0.1\%$ from the target
power production at any timepoint. We also filter the solutions for
non-dominance and therefore remove any solution $\bx$ where there exists
another solution with objective values better than all the objective
values of $\bx$. Finally we sort the solutions by the number of
overloaded transmission lines, total cost, and number of switches in
their corresponding energy network configuration---in that priority. We
then take the weight of the first solution as our weight of choice.

However, even with this a posteriori approach, issues can arise when the
objectives have vastly different scales. For instance, \cref{alg:pf} can
have issues finding appropriate weights when the difference in objective
scales are large and the number of weights \cref{alg:pf} is tasked to
find is low. This results in \cref{alg:pf} requiring a high number of
iterations to assign small weights to objective with large scales to
scale them down.

Therefore, in \cref{sec:energy-norm}, we investigate the effect of
normalizing the objectives in our problem before attempting to find a
set of non-dominated solutions and their corresponding weights with the
method outlined above.

\subsection{Normalization}\label{sec:norm}
We can normalize each {\QUBO} in \cref{eq:fin-obj-true} by using the
fact that we know the solutions $\bx$ that result in the maximum or
minimum objective value for each {\QUBO} matrix in our objective
function. For instance, the cost and overload {\QUBO}
matrices---$\bQ_{C}$ and $\bQ_{O}$---return their maximum values when
the solution sets the all the controllable resources to their maximum
output value, while their minimum objective values are achieved by
setting all controllable resources to their minimum values. The opposite
is true for the power {\QUBO} matrix $\bQ_{P}$. The {\QUBO} matrix
responsible for switching costs $\bQ_{C}$ will be at its maximum
objective value when the solution constantly switches the states of all
controllable resources at every timepoint, while its minimum objective
value state is the solution where no state switching occurs. We then
normalize each {\QUBO} matrix $\bQ$ by dividing it by its respective
range.


\section{Experiments}
\label{sec:experiments}

Our experiments are carried out on redispatch problems constructed using
data from the SimBench dataset~\cite{SimBench2024}. Specifically we use
the \texttt{1-EHV-mixed--0-sw} network which comprises of $338$
controllable resources. Furthermore, we assume that the controllable
resources have $1+4$ states: the \textit{off} state and $4$ linear steps between
the minimum and maximum power each resource is capable of
producing. This results in our problem having
$|\bx| = T \times 338 \times 5 = T\times 1690$ binary decision variables
to optimize over, where $T$ is the number of timepoints considered in
the problem. See \cref{sec:ex-setup} for further details on how we
extract and convert the data from SimBench into the redispatch problems
used in this section.

Based on our problem setup, we investigate various aspects of our model
by conducting four evaluations: First, we study the effectiveness of our
novel normalization for the unbalanced penalization. Second, we
determine meaningful weights for the objectives in our multi-objective
problem, $(\mu_{O}, \mu_{C}, \mu_{S})$, by finding Pareto optimal
solutions on the redispatch problem with $2$ timepoints with classical
solvers. We then evaluate how applicable these weights are when used on
problems with more timepoints. Finally, we compare solutions found by
\cref{alg:a-exp-full} when the subproblems in each iteration are solved
using a classical solver versus a quantum annealer. The code used in
this section can be found at \url{https://gitlab.com/lklee/quantum-power-redispatch}.
\subsection{Unbalanced Penalty Normalization}
\label{sec:ex-unb-pen}
\begin{figure}[t!]
  \centering
  \includegraphics[width=0.8\columnwidth]{./fig/unb\_pen}
  \caption{The 2D histogram over the objective values and number of
    overloaded transmission lines of $10000$ uniformly sampled one-hot
    solutions $\bx$.}
  \label{fig:unb-pen}
\end{figure}
In \cref{sec:cost-overload}, we argued that for unbalanced penalization
to be theoretically sound, the $h_{t,l}(\bx)$ must be normalized such
that $\max_{\bx} h_{t,l}(\bx) \leq 1$. Therefore, in this section we
investigate the effect of unbalanced penalty normalization on the
{\QUBO} objective $\bQ_{O}$ in \cref{eq:cost-overload-ineq}---the
objective that encodes the inequality on the number of overloaded
transmission lines.

To this end, we uniformly sample $10000$ one-hot solution vectors. For
each vector $\bx$, we compute two quantities: the number of overloaded
transmission lines in its corresponding power system configuration, and
its objective value $\bx^{\trans}\bQ_{O}\,\bx$. We perform the latter
computation for both normalized and un-normalized versions of
$\bQ_{O}$. \cref{fig:unb-pen} presents a 2D histogram over these
quantities for the two different versions of $\bQ_{O}$.

From \cref{fig:unb-pen}, we can observe that the objective values for
\textit{normalized} unbalanced penalization increases as the number of
overloaded lines increases; this is desirable as it penalizes solutions
proportionally with the number of overloads.

Conversely, the inverse is true for \textit{un-normalized} unbalanced
penalization. Solutions with few overloaded lines have large
$h_{t,l}(\bx)$ values, which fall outside the region where the
second-order Taylor expansion provides an accurate approximation of the
negative exponential. Consequently, these solutions are penalized
heavily, resulting in the observed negative correlation.

\subsection{Normalization of {\QUBO} Objectives}\label{sec:energy-norm}
\newcommand{\sd}[1]{\scriptsize $\pm #1$}
\begin{table}[!t]
  \caption{Statistics of the non-dominated solutions and properties of
    the top solution found by the approach in \cref{sec:multi-obj}.}
  \label{tab:pf}
  \centering
  \begin{tabular}{l r r r r}
    \toprule
    Statistics of Solutions
    & \multicolumn{2}{c}{Original}
    & \multicolumn{2}{c}{Normalized}\\
    \midrule
    {\# Non-Dominated}
    &\multicolumn{2}{c}{$15$} & \multicolumn{2}{c}{$30$}\\
    {Hypervolume ($10^{14}$)}
    &\multicolumn{2}{c}{$2.7005$}
    &\multicolumn{2}{c}{$3.2669$}\\
    \midrule
    Properties of Best Solution
    & T=1 & T=2 & T=1 & T=2\\
    \midrule
    \# Switches
    &\multicolumn{2}{r}{$194$} & \multicolumn{2}{r}{$156$}\\
    \# Overloads
    &{$7$} &$6$ &{$5$} &$4$\\
    Cost (€$10^{6}$)
    &{$1.1473$} &$1.0320$ &{$1.4452$} &$1.2767$\\
    \bottomrule
  \end{tabular}
\end{table}

\begin{figure*}[t!]
  \centering
  \subfloat[Comparison of solutions found for redispatch
  problems with different timepoints using \cref{alg:a-exp-full} with
  \texttt{TabuSearch} as the subproblem solver. Results are averaged
  over $10$ runs.]{
    \includegraphics[width=0.375\textwidth]{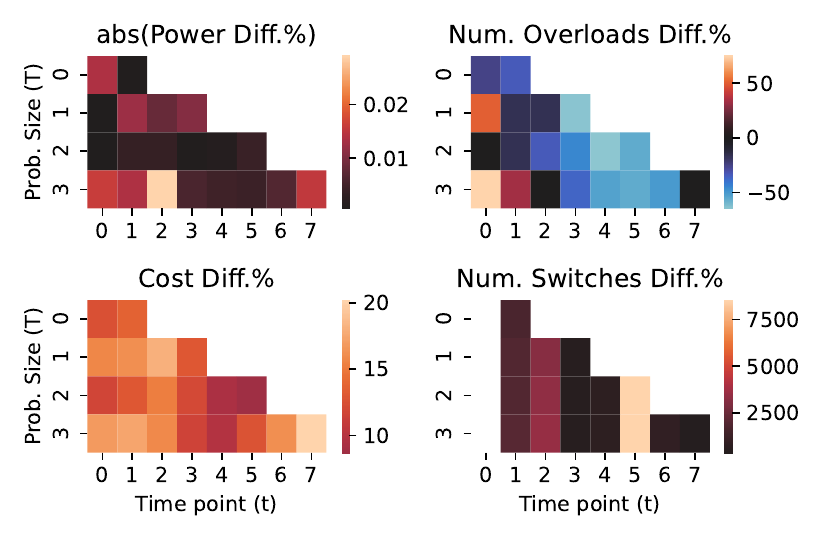}
    \label{fig:apply}
  }
  \hspace{3em}
  \subfloat[Attributes of the solutions for the $T=8$ redispatch problem
  found by \cref{alg:a-exp-full} with either \texttt{TabuSearch} or AQC.
  Plots are over median value for
  each attribute with the bars representing and min/max values.]{
    \includegraphics[width=0.475\textwidth]{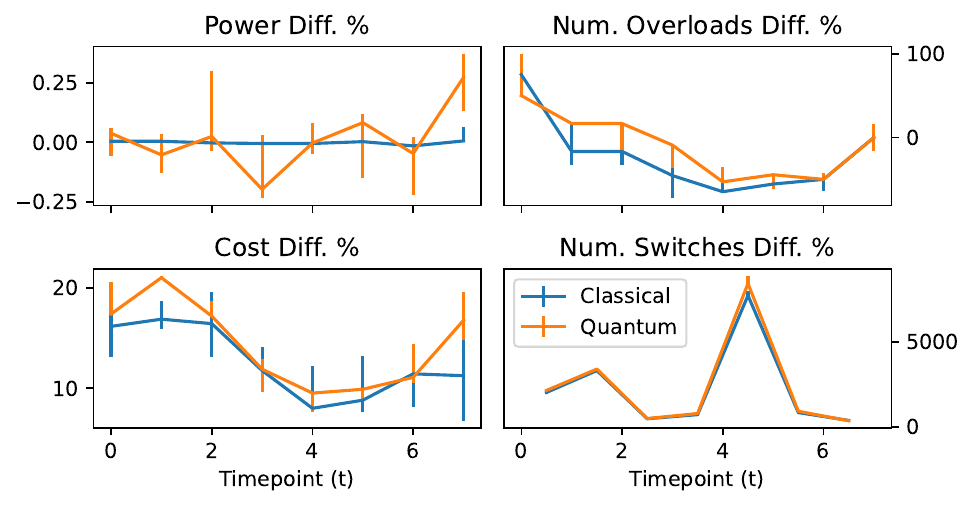}
    \label{fig:compare-solvers}
  }
  \caption{The difference---in percentage---of different attributes
    between the simbech power system configuration and the
    configuration found by solving our redispatch problems. The
    objectives of the redispatch problems are scalarized using weights
    found from \cref{sec:energy-norm} and optimized using
    $\alpha$-expansion (see \cref{alg:a-exp-full}).}
  \label{fig:apply-compare}
\end{figure*}

In this section we assess the effect normalizing our objectives can have
on the non-dominated solutions found by the a posteriori approach
outlined in \cref{sec:multi-obj}. For scalarizations over the
normalized and original {\QUBO} objectives, we use the Lagrangian
multipliers $\lambda_{P}=10^{4}$ and $\lambda_{P}=1$ respectively. We
also instruct \cref{alg:pf} to find $25$ weights, where for each
candidate scalarization, we run the $\alpha$-expansion procedure in
\cref{alg:a-exp-full} $10$ times; therefore we obtain $10$ solutions per
scalarization. We also use the implementation of
\texttt{TabuSearch}~\cite{palubeckis2004} in D-Wave's Ocean SDK as the
subproblem solver in \cref{alg:a-exp-full}. The statistics and
properties of the solutions found by the approach in
\cref{sec:multi-obj} can be found in \cref{tab:pf}.

The hypervolume of a set of non-dominated solutions measures the size of
the objective space---volume in 3D and area in 2D---that is dominated by
at least one of the solutions in the set of non-dominated solutions and
bounded by a reference point~\cite{zitzler1998}; the reference point
used in calculating the hypervolume must be dominated by all the
solutions in the set of non-dominated solutions. Therefore, the larger
the hypervolume, the better. In our case, we compute the hypervolume in
the original \textit{un-normalized} objective space. The reference point
used is then just the maximum objective values for each objective as
described in \cref{sec:norm}. From \cref{tab:pf}, we can observe that
the set of non-dominated solutions found using the normalized objective
has a larger hypervolume than the objectives without
normalization. Therefore, the set of solutions found with the normalized
objectives are more diverse and of higher quality. The a posteriori
approach returned a greater number of non-dominated solutions using the
normalized objectives as well.

Furthermore, the top solution found using the normalized
objectives---compared to the original objectives---has a lower number of
state switches for the controllable resources in the network as well as
a lower number of overloaded transmission lines. However, this comes at
a cost of having a higher production cost overall.

\subsection{Applying Weights to Larger Problems}\label{sec:SimBench}
Recall from \cref{sec:multi-obj} that an issue with a posteriori methods
is that solving multiple scalarizations of our multi-objective functions
can be infeasible for larger problem sizes. Therefore, in this section we
assess the viability of using weights found on redispatch problems
with $2$ timepoints on redispatch problems with $4$, $6$, and $8$
timepoints.

From \cref{fig:apply} we can observe the same pattern from
\cref{sec:energy-norm}---where the number of overloaded transmission
lines in the found configuration is reduced in exchange for a higher
total production cost and number of state switches. Only here, we are
able to observe this pattern across different redispatch problem
sizes. Furthermore, we can observe that at any timepoint for all the
redispatch problems, the total power produced by the found solution has
less than a $\pm{0.02}\%$ deviation from the target power needed to be
produced, thereby ensuring power system stability during operation.

\subsection{Redispatch on Quantum Hardware}\label{sec:compare-solver}
Recall that the original goal of formulating the redispatch problem in
power systems as a {\QUBO} problem is to allow the use of AQC for
finding optimal solutions to the redispatch problem. Therefore, in this
section we demonstrate the ability to use QA---a heuristic form of
AQC---for solving the subproblems during $\alpha$-expansion. Due to
limited compute time on D-Wave's Advantage System 4.1 quantum annealer,
this experiment is done on the $T=8$ redispatch problem over $3$
repetitions, and where $\alpha$-expansion is limited to a maximum of
$20$ iterations. Furthermore, when solving each subproblem via AQC, we
use the solution with the lowest energy out of a sample of $100$.

The attributes of the respective power system configuration for the
solutions found by AQC is in \cref{fig:compare-solvers}; the results
from \texttt{TabuSearch} are also included for reference. From
\cref{fig:compare-solvers}, we can see that the solutions obtained from
using AQC in $\alpha$-expansion does reduce the number of overloaded
transmission lines compared to the power system configuration in
SimBench. However, compared to the configuration found using
\texttt{TabuSearch}, AQC results in a higher overall cost and has a
harder time matching the power production target $\tau$ which can cause
instabilities in the power system.

These results aligns with the current understanding that current day
quantum computers for AQC are still incapable of matching its classical
counterparts due to current quantum computing still being in its 
infancy~\cite{quinton2025}; however, in theory, AQC is capable of a
quadratic speedup over classical simulated
annealing~\cite{mukherjee2015}. Therefore, as the development of 
quantum computers progresses, it is possible for AQC to outperform a
classical approach to solving {\QUBO} problems.

\section{Conclusion}
We proposed a novel, principled \textit{formalization} of the redispatch
problem in the framework of multi-objective quadratic unconstrained
binary optimization.  Due to our generic formalization, this work shall
serve as a starting point for other researchers in the field that want
to address redispatch problems via quantum optimization.

In order to facilitate a proper realization of inequality constraints as
part of the {\QUBO}, we devised a normalized version of the unbalanced
penalty method, by incorporating problem specific knowledge to find the
range of the unbalanced penalty. An empirical evaluation showed that a
\textit{normalized unbalanced penalty} results in an objective function
that is proportional to the number of inequality constraints violated by
a candidate solution $\bx$. We also showed that it is possible to
normalize the objective functions in our current formulation of the
redispatch problem using problem specific knowledge. We then compared
the solutions found by the a posteriori approach in \cref{sec:multi-obj}
utilizing the original and normalized {\QUBO} objective functions; the
solutions obtained using the normalized objective functions were more
diverse and of higher quality.

Moreover, we formulated a novel $\alpha$-Expansion algorithm for
optimizing very large redispatch instances. In particular, we showed how
we can specialize $\alpha$-expansion to not only find solutions that
conform to the one-hot constraint in \cref{eq:con-one-hot}---something
that has been done before~\cite{benkner2021,gerlach2023}---but also
redispatch specific constraints, such as the adjacent state switching
constraint in \cref{eq:con-adj}. We then conducted numerical experiments
on freely available data and showed that our approach can be used to
decompose and solve problem sizes using QA that would have been
otherwise too big for current day quantum annealers. The results
confirmed our theoretical considerations and suggest that our approach
is able to find configurations that result in less overloaded
transmission lines compared to the historical configurations of the same
controllable resources in the German energy network found in the
SimBench dataset.

\section*{Acknowledgment}
This research has partly been funded by the Federal Ministry of
Research, Technology and Space in Germany and the state of North-Rhine
Westphalia as part of the Lamarr-Institute for Machine Learning and
Artificial Intelligence.

\IEEEtriggeratref{15}

\bibliographystyle{IEEEtran}
\bibliography{refs}

\begin{figure*}[t!]
  \centering
  \begin{align*}
    \argmin_{\bx}O(&\bx)\\
    =\argmin_{\bx}
    &\sum_{t=1}^{T}\sum_{l=1}^{L}\Biggl(
       1 - \frac{h_{t,l}(\bx)}{\fnorm{h_{t,l}}}
       + \frac{1}{2}\cdot \frac{h_{t,l}(\bx)^{2}}{\fnorm{h_{t,l}}^{2}}
       \Biggr)\\
    =\argmin_{\bx}
    &\sum_{t=1}^{T}\sum_{l=1}^{L}\frac{-1}{\fnorm{h_{t,l}}}\Biggl(
       [\maxload]_{t,l} - \sum_{a=1}^{n}[\bS]_{a,l}
      \sum_{i=1}^{k}\bx_{t,a,i}\cdot\bp_{a,i}
      \Biggr)\\
    &-\sum_{t=1}^{T}\sum_{l=1}^{L} \frac{1}{\fnorm{h_{t,l}}^{2}} \Biggl(
      [\maxload]_{t,l} \sum_{a=1}^{n}[\bS]_{a,l}
      \sum_{i=1}^{k}\bx_{t,a,i}\cdot\bp_{a,i}
      \Biggr)
      +\sum_{t=1}^{T}\sum_{l=1}^{L} \frac{1}{2\fnorm{h_{t,l}}^{2}}
      \Biggl(\sum_{a=1}^{n}[\bS]_{a,l}
      \sum_{i=1}^{k}\bx_{t,a,i}\cdot\bp_{a,i}
      \Biggr)^{2}\\
    =\argmin_{\bx}
    &\sum_{t=1}^{T} \sum_{a=1}^{n} \sum_{i=1}^{k}
      \bx_{t,a,i}\cdot\bp_{a,i} \sum_{l=1}^{L} [\bS]_{a,l}
      \frac{\fnorm{h_{t,l}} -[\maxload]_{t,l}}{\fnorm{h_{t,l}}^{2}}
    +\sum_{t=1}^{T}
      \sum_{a=1}^{n} \sum_{b=1}^{n} \sum_{i=1}^{k} \sum_{j=1}^{k}
      \bx_{t,a,i}\cdot \bx_{t,b,j}\cdot\bp_{a,i}\cdot\bp_{b,j}
      \sum_{l=1}^{L}
      \frac{[\bS]_{a,l}[\bS]_{b,l}}{2\fnorm{h_{t,l}}^{2}}\\
  \end{align*}
  \caption{Derivation of Overload cost QUBO}
  \label{fig:cost-overload-qubo}
\end{figure*}

\newpage

\appendices

\begin{figure}[b!]
  \centering
  \includegraphics[scale=0.6]{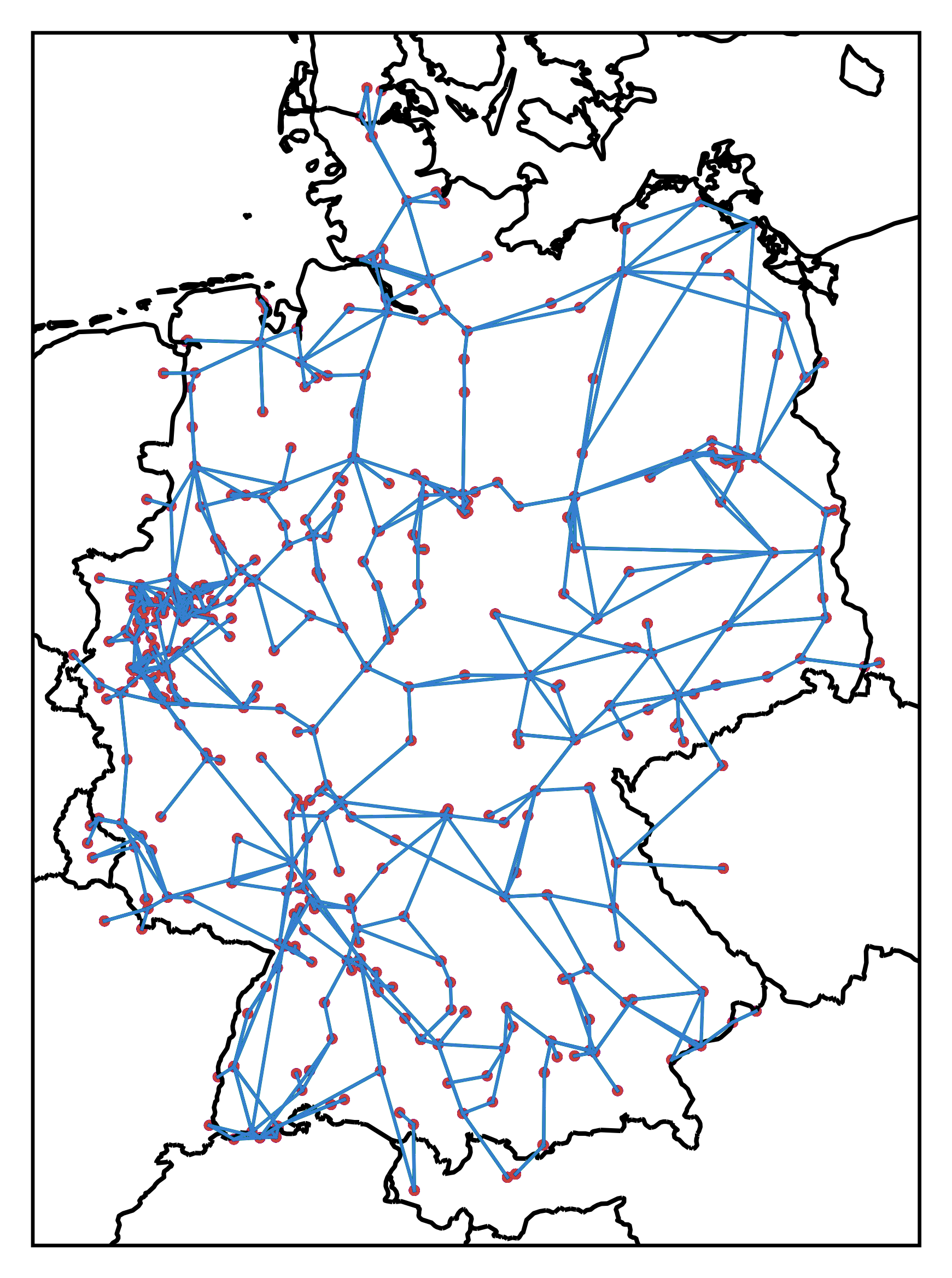}
  \caption{German Power Network of the Extra High
  Voltage Grid in the SimBench dataset.\label{fig:power-network}}
\end{figure}

\section{Experimental Setup}\label[appendix]{sec:ex-setup}
Our problem formulation is only viable if data is available to
instantiate it. In order to facilitate usage of our model and
reproducibility of our results, we provide a walk through for
instantiating the redispatch problems used in \cref{sec:experiments} on
the basis of freely available data.

\subsection{SimBench Data}
We utilize the SimBench dataset~\cite{SimBench2024}---specifically the
\texttt{1-EHV-mixed--0-sw} network---which comprises of 338 controllable
resources. These resources each have a minimum and maximum power output
(MW). The network also includes 225 static generation resources with
fixed power outputs (MW), 390 loads, and 7 external grid connections
that can either inject into or draw power from the power system. These
elements are connected to 3085 buses, interconnected by 849 transmission
lines, each with a maximum current rating. A visualization of the
network's buses and transmission lines is shown in
\cref{fig:power-network}.

\subsection{Extracting data needed from SimBench}
To construct the redispatch problems used in \cref{sec:experiments},
relevant information is first extracted from the SimBench
dataset. Specifically, we extract data from evenly spaced timepoints in
SimBench, where each timepoint is 8 \textit{timesteps} apart and each
\textit{timestep} represents 15 minutes; therefore, the each timepoint is
spaced 2 hours apart from each other. For the \texttt{1-EHV-mixed--0-sw}
network, four element types influence the total power in the system:
controllable resources, static generators, loads, and external grid
connections. The sum over the power produced and consumed by these
elements at each timepoint defines the target power demand $\tau$.

Each controllable resource $a\in\indices{n}$ has a minimum,
$\fmin{\g{\mat{\Phi}}_{\cdot,a}}$, and maximum,
$\fmax{\g{\mat{\Phi}}_{\cdot,a}}$, power output rating. We discretize
the power levels of resource $a$ using linear steps from
$\fmin{\g{\mat{\Phi}}_{\cdot,a}}$ to
$\fmax{\g{\mat{\Phi}}_{\cdot,a}}$. We also define a lowest state,
$\g{\bp}_{a,1}=0$, representing the resource producing 0 MWh of energy.
\begin{equation}
  \label{eq:30}
  \g{\bp}_{a,i} =
  \begin{cases}
    \fmin{\g{\mat{\Phi}}_{\cdot,a}} + \frac{i-1}{k - 1}\cdot
    (\fmax{\g{\mat{\Phi}}_{\cdot,a}} - \fmin{\g{\mat{\Phi}}_{\cdot,a}})
    & \fmin{\g{\mat{\Phi}}_{\cdot,a}} = 0\\
    0 & \fmin{\g{\mat{\Phi}}_{\cdot,a}} \neq 0, i = 1\\
    \fmin{\g{\mat{\Phi}}_{\cdot,a}} + \frac{i-2}{k - 2}\cdot
    (\fmax{\g{\mat{\Phi}}_{\cdot,a}} - \fmin{\g{\mat{\Phi}}_{\cdot,a}})
    & \fmin{\g{\mat{\Phi}}_{\cdot,a}} \neq 0, i > 1
  \end{cases}
\end{equation}
where $k$ is the desired number of states each controllable resource can
be in, and $i\in\indices{k}$. 
\begin{table}[t!]
  \centering
  \begin{tabular}{l r r}
    \toprule
    Resource Type & Min Cost & Max Cost \\
    \midrule
    Hard Coal       & 50 & 90  \\
    Gas             & 40 & 100 \\
    Solar           & 30 & 60  \\
    Nuclear         & 80 & 120 \\
    Offshore wind   & 70 & 120 \\
    Onshore wind    & 40 & 80  \\
    Waste           & 80 & 110 \\
    Lignite         & 40 & 70  \\
    Oil             & 90 & 160 \\
    Imported Energy & 30 & 100 \\
    \bottomrule
  \end{tabular}
  \caption{Minimum and maximum cost in € of generating 1 MWh of energy for each type of Controllable Resources based on levelized cost of electricity.}\label{tbl:cost}
\end{table}

We assume, the production cost of the $a$-th controllable resource at
its $i$-th state can be calculated by multiplying the cost of the
resource producing one MWh of energy, $\g{\bc}_{a}^{*}$, with the amount
of energy the $a$-th controllable resource will produce in the $i$-th
state, $\g{\bp}_{a,i}$:
\begin{equation}
  \label{eq:39}
  \forall t\in\indices{T}:
  \g{\bc}_{t,a,i} \defeq \g{\bc}_{a}^{*} \cdot \g{\bp}_{a,i},
\end{equation}
where $a\in\indices{n}, i\in\indices{k}$. We also assume that the
production cost of all controllable resources does not change over time.

However, the value of $\g{\bc}_{a}^{*}$ depends on the $a$-th
controllable resource's type. In reality, different types of resources
will have different ranges in cost for producing one MWh of energy, as
we can see in \cref{tbl:cost}. To get a concrete cost for our purposes,
we set the value of $\g{\bc}_{a}^{*}$ by sampling from a uniform
distribution between the minimum and maximum cost per MWh,
$\fmins{\g{\bc}_{a}}$ and $\fmaxs{\g{\bc}_{a}}$ respectively, of the
$a$-th controllable resource's type.
\begin{gather}
  \label{eq:33}
  \g{\bc}_{a}^{*} \sim U(\fmins{\g{\bc}_{a}},\fmaxs{\g{\bc}_{a}}).
\end{gather}
Specifically in this paper we use the method \texttt{np.random.uniform}
with a seed of $0$.

The next order of business is to find the sensitivity matrix $\bS$,
which we need to determine how the production of power at some
resource affects the amount of power flowing through each transmission
line. However, before we do that we need to also record the amount
of power produced or consumed by the static resources, loads, and
external grids connected to the power network. This results in a matrix
where $\ld{\mat{\Phi}}_{t,c}$ is the amount of power produced or
consumed by the $c$-th element---letting static resource, load, and
external grid elements share the same index---where negative values in
$\ld{\mat{\Phi}}$ represents power being consumed by the $c$-th
element. We then estimate the sensitivity matrix using the procedure that us described in
\cref{sec:est-sensitivity} of the Appendix with the loss function:
\begin{gather}
  \label{eq:42}
  \ell(\mat{S}) = \| \mat{\Phi}\mat{S} - {\mat{\Psi}}\|_F^2
  \intertext{where:}
  \mat{S} \defeq
  \begin{bmatrix}
    \g{\mat{S}}\\ \ld{\mat{S}}
  \end{bmatrix}
  \qquad
  \mat{\Phi} =
  \begin{bmatrix}
    \g{\mat{\Phi}} & \ld{\mat{\Phi}}
  \end{bmatrix}
\end{gather}
and $\mat{\Psi}$ is a matrix of the amount of power flow over each
transmission line and timepoint.

Finally, we can compute the matrix $\mat{M}$ where $[\mat{M}]_{t,l}$ is
the maximum load the $l$-th transmission line can handle at timepoint
$t$. To do so, we first assume that the pair of substations each line is
connected to transmits electricity at the same voltage $V$. Furthermore,
from SimBench, each transmission line has a maximum rated current it
can transmit $I$. With these simplifications, we arrive at
\begin{gather}
  \label{eq:43}
  \bigl[\mat{M}\bigr]_{t,l} \defeq
  V \cdot I \cdot \sqrt{3} - \bigl[\ld{\mat{\Phi}}\ld{\mat{S}}\bigr]_{t,l}
\end{gather}
which is basically the difference between the maximum load the
transmission line can carry and the amount of power flowing through the
transmission line as a result of the power production and consumption
from the static resources, loads, and external grids attached to the
network.

\section{Estimating the Sensitivity matrix}\label[appendix]{sec:est-sensitivity}
The sensitivity matrix $\mat{S}$ encodes how produced energy is
distributed over the lines of the network.  Here, we explain how
$\mat{S}$ can be obtained from data, e.g., as measured from an actual
power system or obtained from a simulation.  The general procedure
follows~\cite{ospina2023}. However, we consider additional regularity
constraints.  Let
$\ell(\mat{S}) = \|\mat{S} \tilde{\mat{P}} - \tilde{\mat{\tau}}\|_F^2$
denote the loss function of our estimation. $\tilde{\mat{P}}$ and
$\tilde{\mat{\tau}}$ represent the data: $\tilde{\mat{P}}$ denotes the
produced or consumed energy at the controllable and static resources,
$\tilde{\mat{\tau}}$ represents the recorded or simulated loads on the
lines using pandapower~\cite{pandapower2018}. We estimated $\mat{S}$
from data by solving the program
\begin{equation}
  \min_{\mat{S}\in{[0,1]^{m\times n}}} \ell(\mat{S}) \text{ \ \ s.t. \ \ } \min_{(\iota,\kappa)} (\mat{S} \mat{P})_{\iota,\kappa} \geq 0
\end{equation}
via the projected gradient method~\cite{Bertsekas/1999a}. The algorithm
is provided in \cref{alg:pg}.

\begin{algorithm}[!t]
\SetAlgoNoEnd
\SetAlgoVlined
\caption{Projected gradient method for estimating sensitivity matrices.\label{alg:pg}}
\KwIn{$\tilde{\mat{P}}$, $\tilde{\mat{\tau}}$}
\KwOut{$\mat{S}$}

$i \gets 0$

$\mat{S}^0 \gets I_{m,n} $

\While{not converged}{
$\mat{S}^{i + \nicefrac{1}{2}} \leftarrow \mat{S}^i + \eta^i \nabla \ell(\mat{S})  + \eta^i \lambda \mat{P}_{\iota^*,\kappa^*}$

$\mat{S}^{i + 1} \leftarrow \arg\min_{\mat{v} \in{[0,1]^{m\times n}}} \|\mat{v} - \mat{S}^{i + \nicefrac{1}{2}}\|_2^2$

$i \gets i + 1$
}

return $\mat{S}^i$

\end{algorithm}



\end{document}